\documentstyle[12pt] {article}
\newcommand{\beq}{\begin{equation}}
\newcommand{\eeq}{\end{equation}}
\newcommand{\beqa}{\begin{eqnarray}}
\newcommand{\eeqa}{\end{eqnarray}}
\newcommand{\ba}{\begin{array}}
\newcommand{\ea}{\end{array}}

\begin{document}

\begin{center}
\large
{\bf A Note on the Toda Criterion for\\
Interacting Dipole--Quadrupole Vibrations} 

\vskip 1. truecm

\normalsize
{\bf V.R. Manfredi}$^{(a)(b)}$\footnote{e-mail: manfredi@padova.infn.it} 
and {\bf L. Salasnich}$^{(b)(c)}$\footnote{e-mail: salasnich@math.unipd.it} 
\vskip 0.5 truecm
$^{(a)}$ Dipartimento di Fisica "G. Galilei" \\
Universit\`a di Padova, Via Marzolo 8, I--35131 Padova, Italy \\
$^{(b)}$ Istituto Nazionale di Fisica Nucleare, Sezione di Padova \\
Via Marzolo 8, I--35131 Padova, Italy \\ 
$^{(c)}$Dipartimento di Matematica Pura ed Applicata \\
Universit\`a di Padova, Via Belzoni 7, I--35131 Padova, Italy \\
Istituto Nazionale di Fisica della Materia, Unit\`a di Milano \\
Via Celoria 16, I--20133 Milano, Italy
\end{center}

\vskip 0.5 truecm

\begin{center}
{\bf Abstract}
\end{center}
\par
The Toda criterion of the Gaussian curvature is applied 
to calculate analytically the transition energy 
from regular to chaotic motion in a schematic model describing 
the interaction between collective dipole and quadrupole modes in atomic 
nuclei. 

\newpage

\par
In recent years great interest has been shown in nonlinear effects in 
nuclear$^{1)}$ and particle physics$^{2)}$. These phenomena may be 
considered the precursors of order--chaos transition. 
In these systems this transition has usually been 
studied numerically with Lyapunov exponents and Poincar\`e sections$^{3)}$. 
Less attention has been paid to analytical criteria. 
\par
In this paper we study the hamiltonian: 
\beq
H={1\over 2}(B_1 {\dot q_1}^2 + C_1 q_1^2) + 
{1\over 2}(B_2 {\dot q_2}^2 + C_2 q_2^2) 
- k_1 q_1^2 q_2 \; ,
\eeq
introduced by Bohr and Mottelson$^{4)}$ in order to describe the coupling 
between the giant--dipole resonance (GDR) and the quadrupole excitation, 
later used by the authors of reference$^{5)}$. 
The vibration in the $q_1$ direction is associated 
with the GDR, while the $q_2$ motion plays the role of 
quadrupole deformation of the nuclear surface. Here the parameters of 
inertia $B_i$ and of stiffness $C_i$ ($i=1,2$) are real constants. 
The perturbing potential $k_1q_1^2q_2$ 
induces a splitting of the dipole 
frequency caused by the static quadrupole deformation. 
Bohr and Mottelson$^{4)}$ give the estimation $k_1=(5/2)\sqrt{3/(2\pi )}C_1$ 
for the coupling parameter, in order 
to reproduce the experimental data of fotoabsorption in deformed nuclei$^{4)}$. 
\par 
The aim of this work is to apply the Toda criterion of the Gaussian 
Curvature$^{6)}$ to the 
Hamiltonian (1) in order to calculate the energy of the onset of 
chaotic motion, henceforth $E_c$. 
\par
The Toda criterion is based on a local estimation of the rate of separation 
of neighboring trajectories in the classical phase space of the 
model$^{6),7)}$. 
To obtain the time evolution of a dynamical system with the Hamiltonian 
\beq
H={p_1^2\over 2B_1}+{p_2^2\over 2B_2} + V(q_1, q_2 ) \; , 
\eeq
where $p_k=B_k {\dot q_k}$, $k=1,2$, 
the following equations have to be solved: 
\beq
{d{\bf q}\over dt}={\partial H\over \partial {\bf p} } \; 
, \;\;\;\;\;\; 
{d{\bf p}\over dt}= - {\partial H\over \partial {\bf q}} \; ,
\eeq
where ${\bf q}=(q_1,q_2)$ and ${\bf p}= (p_1,p_2)$. 
The linearized equation of motion for the deviations are
\beq
{d \delta {\bf p} \over dt} = M^{-1} \delta {\bf p} \; , 
\;\;\;\;\;\; 
{d \delta {\bf q}\over dt} = - S(t) \delta {\bf q} \; , 
\eeq
where $M_{ij}^{-1}=\delta_{ij}B_i^{-1}$, and 
\beq
S_{ij}(t) = {\partial^2 V\over \partial q_i q_j}|_{{\bf q}={\bf q}(t)} 
\; ,
\eeq
where ${\bf q}(t)$ is the solution of (3). 
The stability of the dynamical system is then determined by the 
eigenvalues of the $4\times 4$ stability matrix
\beq
\Gamma({\bf q}(t))= 
\left( \begin{array}{cc}  
            0     &   M^{-1} \\ 
         -S(t)    &   0 
\end{array}  \right) \; .
\eeq
If at least one of the eigenvalues $\lambda_i$ 
of the stability matrix $\Gamma$ 
is real, then the separation of the trajectories grows exponentially 
and the motion is unstable. 
Imaginary eigenvalues correspond to stable motion. 
\par
To diagonalize the matrix $\Gamma $, we must first solve the equation of motion 
(3). The problem can be significantly simplified by assuming that the 
time dependence can be eliminated by replacement of the time--dependent 
point ${\bf q}(t)$ 
of configuration space by a time--independent coordinate ${\bf q}$, 
i.e. $\Gamma ({\bf q}(t))=\Gamma ({\bf q})$. 
The eigenvalues then are
\beq
\lambda = \pm [-b\pm \sqrt{b^2 - 4 c} ]^{1\over 2} \; ,
\eeq
where 
\beq
b = B_1^{-1}B_2^{-1} 
\big[ {\partial^2 V\over \partial q_1^2} + 
 {\partial^2 V\over \partial q_2^2} \big] \; ,
\eeq
\beq
c = B_1^{-1}B_2^{-1} \big[ {\partial^2 V\over \partial q_1^2} 
{\partial^2 V\over \partial q_2^2} - 
({\partial^2 V\over \partial q_1 q_2})^2 \big] \; .
\eeq
Now, if $b>0$ then with $c\geq 0$ the eigenvalues are purely imaginary 
and the motion is stable, while with $c<0$ the pair of eigenvalues 
becomes real and this leads to exponential separation of neighboring 
trajectories, i.e. chaotic motion. The parameter $c$ has the same sign 
as the Gaussian curvature $K_G$ of the potential--energy surface: 
\beq
K_G(q_1,q_2) = { {\partial^2 V\over \partial q_1^2} 
{\partial^2 V\over \partial q_2^2} - 
({\partial^2 V\over \partial q_1 q_2})^2 \over 
\big[ 1+ ({\partial^2 V\over \partial q_1^2})^2 + 
({\partial^2 V\over \partial q_2^2})^2 \big]^2 } \; . 
\eeq
For our nuclear model the potential energy is given by 
\beq
V(q_1 ,q_2)={1\over 2}(C_1q_1^2 + C_2q_2^2) 
-k_1 q_1^2 q_2 \; ,
\eeq
and the $\Gamma$ matrix reads
\beq
\Gamma (q_1,q_2) =
\left( \begin{array}{cccc}  
         0         &    0         & B_1{-1} & 0         \\ 
         0         &    0         & 0       & B_2^{-1}  \\ 
 -C_1+2k_1q_2 & 2k_1q_1 & 0       & 0         \\ 
  2k_1q_1     & -C_2         & 0       & 0  
\end{array}  \right) 
\eeq
The potential has one minimum for $q_1=q_2 =0$ with $V=0$, 
and two saddle points for $q_1 = \pm \sqrt{C_1C_2\over 2k_1^2}$, 
$q_2 = {C_1^2 C_2 \over 8 k_1^2}$. 
These are the points for which the curvature criterion is exact 
(they are the fixed points of the Hamilton equations). The origin 
is a stable elliptic point and the two saddle points are unstable 
hyperbolic points. 
\par
The dissociation energy of this model is $E_D={C_1^2C_2\over 8k_1^2}$ 
and it occurs at the 
saddle points of the potential energy surface. 
Classically, a particle located within the potential well 
with energy $0<E<E_D$ is always bound. 
\par
The Gaussian curvature vanishes at the points that satisfy the 
equation
\beq
C_2(C_1 - 2k_1 q_2 ) - 4 k_1^2 q_1^2 = 0 \; ,
\eeq
where 
\beq
q_1^2 = {C_2\over 4k_1^2} (C_1 -2 k_1 q_2) \; ,
\eeq
with $q_2\leq {C_1\over 2k_1}$. 
At low energies, the motion near the minimum of the potential 
energy, where the curvature is positive, is periodic or quasi--periodic 
and is separated from the region of instability by a line of zero 
curvature; if the energy increases, the system will be, for some initial 
conditions, in a region of negative curvature where the motion is chaotic. 
In accordance with this scenario, the energy of order$\to$chaos 
transition $E_c$ is equal to the minimum value of the line of 
zero Gaussian curvature $K_G$ of the potential--energy surface of the 
system$^{6),7)}$. The energy of the zero--curvature line is determined 
by the expression 
\beq
V(K_G=0,q_2)=C_2q_2^2 - {C_1C_2\over 2k_1}q_2 + 
{C_1^2 C_2\over 8 k_1^2} \; .
\eeq
It is easy to show that the minimal energy on the zero--curvature line 
is given by 
\beq
E_c = V(K_G=0, q_2 = {C_1\over 4k_1}) = {C_1^2 C_2 \over 16  k_1^2} \; .
\eeq
This is the energy of the onset of chaos of the 
model and we see that $E_c=E_D/2$. 
\par
With the estimation $k_1=(5/2)\sqrt{2/(2\pi )}C_1$ we have
\beq
E_c = {\pi \over 150} C_2 \; .
\eeq
By using the charged liquid drop approximation$^{4)}$ for the stiffness 
coefficient $C_2$, we obtain for $E_c$ the following relation 
\beq
E_c = \big( 
{b_s\over \pi }A^{2\over 3} -{3\over 10 \pi }{Z^2 e^2\over r_0 A^{1/3}} \big) 
{\pi \over 150} \; ,
\eeq
where $b_s\simeq 17$ MeV, $r_0\simeq 1.2$ fm, 
$A$ and $Z$ are the atomic and proton 
numbers respectively (see Figure 1). 
With the liquid--drop approximation$^{4)}$, 
the Coulomb repulsion counteracts the effect of the surface tension 
and leads to negative values of $C_2$ for $Z^2/A > (10/3)(b_s r_0/e^2) 
\simeq 49$. 
\par
We can use the experimental quadrupole frequency $\hbar \omega_2$ 
and $B(E_2)$ values to determine$^{4)}$ the restoring force parameter 
$C_2$. In this case 
\beq
E_c = {\pi \over 60} \hbar \omega_2 
({3\over 4\pi} Ze R_0^2 )^2 B(E2;0\to 2)^{-1} \; .
\eeq
The experimental values of $C_2$ exhibit variations greater by a 
factor $10$ above and below the liquid--drop estimation. 
This effect can be understood in terms of the approach to instability 
as particles are added to the closed shells and so 
$E_c$ is in the range $\simeq 0.5$--$5$ MeV depending 
on the nucleus. 
\par
The curvature criterion is able to 
characterize the local behaviour of the system (for example the local 
instability) and may give only a signature of the global 
properties (e.g. the global instability). As is well known, a very 
useful tool for the study of global properties is provided by the 
Poincar\`e sections$^{9)}$. With this aim the classical trajectories have 
been calculated by the fourth order Runge--Kutta method. In order to avoid 
numerical errors connected to the use of finite temporal intervals, a 
first--order interpolation has been used$^{10)}$.
\par
The Hamilton equations of the systems are:
$$
{\dot q_1}=B_1 p_1 \; , \;\;\;\;
{\dot q_2}=B_2 p_2 \; ,
$$
\beq
{\dot p_1}=-C_1 q_1 + 2 k_1 q_1 q_2 \; , \;\;\;\;
{\dot p_2}=-2C_2 q_2 + k_1 q_1^2 \; .
\eeq
Figure 2 shows the Poincar\`e sections for different values 
of the energy. Below the critical energy $E_c$ the system is regular but 
above $E_c$, the Poincar\`e sections
clearly show an order--chaos transition as the energy 
increases. 
\par
In conclusion, the Poincar\`e sections show that 
in our schematic model, which describes 
the interaction between collective dipole and quadrupole modes in atomic 
nuclei, the Toda criterion of the Gaussian curvature gives 
a quite good estimate of the energy $E_c$ of the onset of chaos. 
This critical energy $E_c$ depends on the stiffness coefficient 
of the quadrupole vibration, which 
can be estimated analytically within the liquid--drop approximation 
or can be experimentally determined by using the quadrupole frequency 
$\hbar \omega_2$ and the $B(E_2)$ transition probabilities. 

\begin{center}
*****
\end{center}
\par
LS is greatly indebted to Prof. M. Robnik and Dr. A. Rapisarda 
for many enlightening discussions. 

\newpage

\section*{References}

\begin{description}

\item{\ 1.} M.T. Lopez--Arias, V.R. Manfredi 
and L. Salasnich, Riv. Nuovo Cimento, 
{\bf 17}, N. 5 (1994); V.R. Manfredi and L. Salasnich: 
Int. J. Mod. Phys. {\bf E 4}, 625 (1995). 

\item{\ 2.} T. Kawabe, Phys. Lett. {\bf B 343}, 254 (1995); 
L. Salasnich, {\it Phys. Rev.} {\bf D 52}, 6189 (1995); 
L. Salasnich, Mod. Phys. Lett. {\bf A 10}, 3119 (1995). 

\item{\ 3.} V.R. Manfredi, M. Rosa--Clot, L. Salasnich and S. Taddei, 
Int. J. Mod. Phys. {\bf E 5}, 519 (1996). 

\item{\ 4.} A. Bohr and B.R. Mottelson, {\it Nuclear Structure}, vol. 2 
(Benjamin, New York, 1975). 

\item{\ 5.} W.E. Ormand, M. Borromeo, R.A. Broglia, G. Lazzari, T. Dossing and 
B. Lauritzen, Phys. Rev. {\bf A 47}, 4556 (1989). 

\item{\ 6.} M. Toda, Phys. Lett. {\bf A 48}, 335 (1974); 
G. Benettin, R. Brambilla and L. Galgani, Physica {\bf A 87}, 381 (1977). 

\item{\ 7.} Yu.L. Bolotin, V.Yu. Gonchar, F.V. Inopin, V.V. Levenko, 
V.N. Tartasov and N.A. Chekanov, J. Sov. Part. Nucl. {\bf 20}, 372 (1989). 

\item{\ 8.} V.R. Manfredi and L. Salasnich, 
Nuovo Cimento {\bf A 108}, 449 (1995). 

\item{\ 9.} H. Poincar\`e, {\it New Methods of Celestial Mechanics}, 
vol. 3, ch. 27 (Transl. NASA Washington DC 1967).

\item{\ 10.} M. Henon, Physica {\bf D 5}, 412 (1982).

\end{description}

\newpage

\section*{Figure Captions}
\vskip 0.5 truecm

\parindent=0.pt

{\bf Figure 1}: The critical energy $E_c$ {\it versus} the atomic number $A$,  
with $Z=A/2$. 

{\bf Figure 2}: The Poincar\`e sections 
for different values of the energy; from the top--left, clockwise: 
$E=0.02$, $E=0.03$, $E=0.035$ and $E=0.04$. We chose $B_1=B_2=1$, 
$C_1=100$ and $C_2=1$. The critical energy is 
$E_c = {\pi \over 150} C_2 \simeq 0.021$. 

\end{document}